\documentstyle[amssymb,aps,twocolumn]{revtex}

\begin{document}
\title{Quantum control of two interacting electrons in a coupled quantum dot}
\author{Ping Zhang$^{1,2}$, Xian-Geng Zhao$^2$}
\address{$^1$International Center of Quantum Structure and State Key laboratory for\\
surface Physics, Institute of Physics, The Chinese Academy of Sciences,
Beijing 100080, P.R. China\\
$^2$Institute of Applied Physics and Computational Mathematics, Beijing\\
100080, P.R. China}
\maketitle

\begin{abstract}
Quantum-state engineering, i.e., active manipulation over the coherent
dynamics of suitable quantum-mechanical systems, has become a fascinating
prospect of modern physics. Here we discuss the dynamics of two interacting
electrons in a coupled quantum dot driven by external electric field. We
show the two quantum dots can be used to prepare maximally entangled Bell
state by varying the strength and duration of an oscillatory electric field.
Different from suggestion given by Loss {\it et al}.[Phys. Rev. A, {\bf 57}
(1998) 120], the present entanglement involves the spatial degree of freedom
for the two electrons. We also find that the coherent tunneling suppression
discussed by Grossmann {\it et al}.[Phys. Rev. Lett., {\bf 67} (1991) 516]
persists in the two-particle case, i.e., two electrons initially localized
in one dot can remain dynamically localized, although the strong Coulomb
repulsion prevents them behaving so. Surprisingly, the interaction enhances
the degree of localization to a larger extent compared to non-interacting
case. We can call this phenomenon Coulomb-enhanced dynamical localization.%
\newline%
\newline%
\end{abstract}

\section{Introduction}

Coherent control of quantum systems has been attracted considerable
attentions in recent years. A basic ingredient of quantum control is
field-induced localization of a single electron in a double-trap\cite
{Grifoni}. The initial efforts devoted to acquiring conditions to maintain
existing localization with oscillatory electric field, and to create and
maintain localization with a semi-infinite oscillatory field\cite
{Gross1,Bavli}. It was found that the investigation of localization can be
approximated by a two-state model consisting of the lowest symmetric and
antisymmetric states of the double-trap potential\cite{Gomez,Gross}. In this
case, perfect localization can be achieved with a strong time-periodic
electric field that causes the Floquet quasienergies to be degenerate.
Later, localization in superlattice systems\cite{Hone}, in dissipative
environments\cite{Cuk}, in molecular systems\cite{Zuo}, induced by
ultrashort laser pulses, and in trapped Bose-Einstein condensates\cite
{Pu,ping}, by means of oscillatory magnetic fields, has been studied.

When two or more interacting particles are present, apart from the highly
nontrivial problem of whether the strong many-body interaction can be
overcome for the particles to create and preserve localization, the
possibility of entanglement of the many-body wave functions arises\cite
{Ping2,Metiu}. Entanglement is an essential ingredient in any scheme of
quantum information processing like quantum information cryptography and
quantum computation, and therefore it is a problem of great current interest
to find or design systems where entanglement can be manipulated\cite{Steane}%
. Most of the theoretical and experimental activity until now has been
associated with atomic and quantum-optic systems. Two- \cite{Tur,Hagley},
three-\cite{Rau}, and four-particle \cite{Sack} entanglement have been
successfully demonstrated experimentally in trapped ions, Rydberg atoms, and
cavity QED. However, a further increase of the number of entangled particles
in these systems is expected to be a severe experimental challenge.
Recently, solid-state realizations of the entanglement have received
increasingly attention due to the fact that semiconductor nanostructures
such as quantum dots (QDs) and double quantum dots (DQDs) with well-defined
atom-like and molecule-like properties, have been fabricated and studied by
many groups \cite{Oo,To}. Kane \cite{Kane} has proposed a scheme which
encodes information onto the nuclear spins of donor atoms in doped silicon
electronic devices where externally applied electric fields are used to
perform logical operations on individual spins. Loss and DiVincenzo\cite
{Loss} have presented a scheme based on spin exchange interaction effects.
More recently, Imamoglu {\it et al}. \cite{Imamoglu} have considered a
quantum computer model based on both electron spins and cavity QED which is
capable of realizing controlled interactions between two distant quantum
dots. Quiroga and Johnson \cite{Quiroga} have suggested that the resonant
transfer interaction between spatially separated excitons in quantum dots
can be exploited to produce maximally entangled Bell states.

In the present work we study the coherent control of the quantum system
consisting of two interacting electrons in a coupled quantum dot (see Fig.
1). With the initial state chosen to be in the spin-singlet space, the
dynamics is reduced to be confined to a three-dimensional Hilbert space, in
which the three basis vectors are equivalent to the eigenstates of $\widehat{%
z}$ component of spin-1 operator. We show that the maximally entangled Bell
state can be prepared and maintained with a pulse of oscillatory electric
field. Also we find that although the Coulomb repulsion between the two
electrons is very strong, dynamical localization can fully build up in the
system parameter manifold which corresponds to the exact crossing of the
quasienergies developed from the unperturbed nearly-degenerate levels.

\section{The model}

The Hamiltonian which we use to describe the dynamics of two interacting
electrons in a coupled quantum dot driven by electric fields is

\begin{equation}
H(t)=\sum_{i=1,2}h({\bf r}_i,{\bf p}_i,t)+C,  \eqnum{1a}
\end{equation}
\begin{equation}
h({\bf r},{\bf p},t)=\frac{{\bf p}^2}{2m}-ezF(t)+V_l({\bf r})+V_v({\bf r}), 
\eqnum{1b}
\end{equation}
\begin{equation}
C=\frac{e^2}{\kappa |{\bf r}_1-{\bf r}_2|},  \eqnum{1c}
\end{equation}
where $C$ is the Coulomb interaction and $h$ the single-particle
Hamiltonian. The dielectric constant $\kappa $ and the effective mass $m$
are material parameters. The potential $V_l$ in $h$ describes the lateral
confinement, whereas $V_v$ models the vertical double-well structure. The
lateral coupling of the dots is modeled by a quartic potential 
\begin{equation}
V_l(x,y)=\frac{m\omega _z^2}2\alpha ^2(x^2+y^2),  \eqnum{2}
\end{equation}
where we have introduced the isotropy parameter $\alpha $ determining the
strength of the vertical relative to the lateral confinement. The lateral
effective Bohr radius $a_{B//}=\sqrt{\hslash /(m\omega _z\alpha )}$ is a
measure for the lateral extension of the electron wave function in the dots.
It has been shown in experiments with electrically gated quantum dots in a
two-dimensional electron system that the electronic spectrum is well
described by a simple harmonic oscillator\cite{Jacak}. In describing the
confinement $V_v$ along the inter-dot axis, we have used a (locally
harmonic) double well potential of the form

\begin{equation}
V_v=\frac{m\omega _z^2}{8a^2}(z^2-a^2)^2,  \eqnum{3}
\end{equation}
which separates into two harmonic wells of frequency $\omega _z$(one for
each dot) in the limit $a\gg a_{B\perp }$, where $a$ is half the distance
between the dots and $a_{B\bot }=\sqrt{\hslash /m\omega _z}$ is the vertical
effective Bohr radius. Although in principle a square-well potential would
be a more accurate description of the real potential than the harmonic
double well, we believe there is no qualitative difference between the
results presented below obtained with harmonic potential and the
corresponding results using square-well potential. Finally, the
time-dependent electric field has a DC-AC form, i.e., $F(t)=$ $F_0+F_1\sin
\omega t$.

\section{Spin-1 representation of the Hamiltonian}

Following Burkard {\it et al}.[22] we employ the Hund-Mulliken method of
molecular orbits to describe the low-lying spectrum of our system. This
approach accounts for double occupancies and is therefore suitable for
investigating the questions at issue here. As a starting point for our
calculations we consider the problem of an electron in a single quantum dot.
The one-particle Hamiltonian by which we describe a single electron in the
right (left) dot of the double-dot system is 
\begin{equation}
h_{\pm a}^0({\bf r})=\frac{{\bf p}^2}{2m}+\frac{m\omega _z^2}2\left( \alpha
^2(x^2+y^2)+(z\mp a)^2\right) ,  \eqnum{4}
\end{equation}
and has the ground-state solution

\begin{equation}
\varphi _{\pm a}({\bf r})=\left( \frac{m\omega _z}{\pi \hslash }\right)
^{3/4}\sqrt{\alpha }\exp \left( -\frac{m\omega _z}{2\hslash }\left( \alpha
^2\left( x^2+y^2\right) \right) +\left( z\mp a\right) ^2\right) ,  \eqnum{5}
\end{equation}
corresponding to ground-state energy $\epsilon _{\pm }=\hslash \omega
_z(1+2\alpha )/2$. The two local ground states are not orthogonal and their
overlap is

\begin{equation}
S=\int d^3r\varphi _{+a}^{*}({\bf r})\varphi _{-a}({\bf r})=\exp (-d^2), 
\eqnum{6}
\end{equation}
where the dimensionless parameter $d=a/a_{B\bot }$ denotes the distance
between the two dots. A nonvanishing overlap $S$ implies that the electrons
can tunnel between the dots. From these non-orthogonal states, we construct
the orthonormalized one-particle wave functions 
\begin{equation}
\Phi _{+a}({\bf r})=\frac 1{\sqrt{1-2Sg-g^2}}(\varphi _{+a}-g\varphi _{-a}),
\eqnum{7a}
\end{equation}
\begin{equation}
\Phi _{-a}({\bf r})=\frac 1{\sqrt{1-2Sg-g^2}}(\varphi _{-a}-g\varphi _{+a}),
\eqnum{7b}
\end{equation}
with $g=(1-\sqrt{1-S^2})/S$. For appropriate values of system parameters
such as the inter-dot distance, the overlap $S$ becomes exponentially small
as given in Eq. (7). In this limit an electron in one of the states $\Phi
_{+a}$, $\Phi _{-a}$ is predominately localized around $(x,y,\pm a)$. In the
following we consider this case and use these states to define the localized
single-particle state. Schliemann {\it et al.}\cite{Schliemann}{\it \ }have
used theses two local states as qubits, which are realized by the spin state
of an electron in either orbital $\Phi _{+a}$, or orbital $\Phi _{-a}$.

Using $\Phi _{\pm a}$ we generate six basis functions with respect to which
we diagonalize the two-particle Hamiltonian $H$: The states with double
occupation, $\Psi _{\mp a}^d({\bf r}_1,{\bf r}_2)=\Phi _{\mp a}({\bf r}%
_1)\Phi _{\mp a}({\bf r}_2)$ [denoted by (1,0,0,0)$^T$ and (0,1,0,0)$^T$,
respectively] and the states with single occupation $\Psi _{\pm }^s({\bf r}%
_1,{\bf r}_2)=[\Phi _{+a}({\bf r}_1)\Phi _{-a}({\bf r}_2)\pm \Phi _{-a}({\bf %
r}_1)\Phi _{+a}({\bf r}_2)]$ [denoted by (0,0,1,0)$^T$ and (0,0,0,1)$^T$,
respectively]. Calculating the matrix elements of the Hamiltonian $H(t)$ in
this orthonormal basis we get 
\begin{equation}
H(t)=\left( 
\begin{array}{ll}
H_1(t) & 0 \\ 
0 & (2\epsilon +U_3)I_{1\times 1}
\end{array}
\right) =(2\epsilon +U_3)I_{1\times 1}\oplus H_1(t),  \eqnum{8}
\end{equation}
where $I_{1\times 1}$ is a $1\times 1$ unit matrix, and $H_1$ is 
\begin{equation}
H_1(t)=\left( 
\begin{array}{lll}
2\epsilon +U_1+\mu (t) & \sqrt{2}w & v \\ 
\sqrt{2}w & 2\epsilon +U_2 & \sqrt{2}w \\ 
v & \sqrt{2}w & 2\epsilon +U_1-\mu (t)
\end{array}
\right)  \eqnum{9}
\end{equation}
with the matrix elements given by

\begin{equation}
\epsilon =\langle \Phi _{\pm a}|h(z\mp a)|\Phi _{\pm a}\rangle ,U_1=\langle
\Psi _{\pm a}^d|C|\Psi _{\pm a}^d\rangle ,  \eqnum{10a}
\end{equation}
\begin{equation}
U_2=\langle \Psi _{+}^s|C|\Psi _{+}^s\rangle ,U_3=\langle \Psi _{-}^s|C|\Psi
_{-}^s\rangle ,  \eqnum{10b}
\end{equation}
\begin{equation}
w=\langle \Phi _{\pm a}|h(z\mp a)|\Phi _{\mp a}\rangle +\langle \Psi
_{+}^s|C|\Psi _{\pm a}^d\rangle ,v=\langle \Psi _{+a}^d|C|\Psi
_{-a}^d\rangle ,  \eqnum{10c}
\end{equation}
\begin{equation}
\mu (t)=\mu _0+\mu _1\sin (\omega t)=\frac{2ea(1-g^2)}{1-2Sg+g^2}%
(F_0+F_1\sin \omega t).  \eqnum{10d}
\end{equation}
Clearly, $U_1$ denotes the intra-dot Coulomb interaction, whereas $U_2$ and $%
U_3$ describe the inter-dot Coulomb interaction, without particle transfer. $%
w$ describes single-particle tunneling induced by dot-dot coupling and
Coulomb interaction. $\mu _i=2eF_ia(1-g^2)/(1-2Sg+g^2)$ ($i=0,1$) describes
the coupling strength between the electrons and external electric fields.

Obviously the spin-triplet state $\Psi _{-}^s$ is an eigenstate of the
Hamiltonian (8) with the electron number on each quantum dot invariably one
and has no response to the presence of the electric fields. This is one
important fact involving spin-triplet space. Hence we will focus our
attention on the reduced spin-singlet Hamiltonian $H_1(t)$. Furthermore, the
matrix element $v$ is the amplitude of simultaneous transfer of two
electrons and therefore denotes cotunneling process. Compared to the
single-electron tunneling term $w$, this term can be negligible. Thus the
reduced Hamiltonian can be conveniently rewritten in terms of spin-1
operators

\begin{equation}
H_1(t)=(2\epsilon +U_2)-\mu (t)J_z+uJ_z^2+2wJ_x,  \eqnum{11}
\end{equation}
where $u=U_1-U_2$ is the effective Coulomb interaction, and $J_i$ ($i=x,y,z$%
) are spin-1 operators defined as

\begin{equation}
J_x=\frac 1{\sqrt{2}}\left( 
\begin{array}{ccc}
0 & 1 & 0 \\ 
1 & 0 & 1 \\ 
0 & 1 & 0
\end{array}
\right) ,  \eqnum{12a}
\end{equation}
\begin{equation}
J_y=\frac i{\sqrt{2}}\left( 
\begin{array}{ccc}
0 & -1 & 0 \\ 
1 & 0 & -1 \\ 
0 & 1 & 0
\end{array}
\right) ,  \eqnum{12b}
\end{equation}
and 
\begin{equation}
J_z=\left( 
\begin{array}{ccc}
1 & 0 & 0 \\ 
0 & 0 & 0 \\ 
0 & 0 & -1
\end{array}
\right) .  \eqnum{12c}
\end{equation}
Therefore, the localized two-particle state $\Psi _{-a}^d$ is equivalent to
the eigenstate $\mid j_z=1\rangle $ of $J_z$ and $\Psi _{+a}^d$ to the state 
$\mid j_z=-1\rangle $, whereas the delocalized state $\Psi _{+}^s$ is
identical to the state $\mid j_z=0\rangle $. In the following we will denote 
$\Psi _{-a}^d$ with $\mid LL\rangle $, $\Psi _{+a}^d$ with $\mid RR\rangle $%
, and $\Psi _{+}^s$ with $\mid LR\rangle $.

The first term in Eq. (11) denotes a constant energy shift, and will be
neglected in the following discussions. The evolution of any initial state $%
\mid \Psi (0)\rangle $ under the action of $H_1$ in Eq. (11) can be
expressed as $\mid \Psi (t)\rangle =C_1(t)\mid LL\rangle +C_2(t)\mid
LR\rangle +C_3(t)\mid RR\rangle $ where the coefficients $C_k(t)$ are
determined by the time dependent Schr\"{o}dinger equation (In the following
we set $\hslash =1$) 
\begin{equation}
i\frac d{dt}\left( 
\begin{array}{l}
C_1 \\ 
C_2 \\ 
C_3
\end{array}
\right) =\left( 
\begin{array}{lll}
u+\mu (t) & \sqrt{2}w & 0 \\ 
\sqrt{2}w & 0 & \sqrt{2}w \\ 
0 & \sqrt{2}w & u-\mu (t)
\end{array}
\right) \left( 
\begin{array}{l}
C_1 \\ 
C_2 \\ 
C_3
\end{array}
\right) ,  \eqnum{13}
\end{equation}
and the chosen initial condition $\mid \Psi (0)\rangle $. In the absence of
external electric fields, the eigenenergies of $H_1$ can be easily solved as 
\begin{equation}
E_1=\frac 12\left( u-\sqrt{u^2+16w^2}\right) ,  \eqnum{14a}
\end{equation}
\begin{equation}
E_2=u,  \eqnum{14b}
\end{equation}
\begin{equation}
E_3=\frac 12\left( u+\sqrt{u^2+16w^2}\right) ,  \eqnum{14c}
\end{equation}
and the corresponding eigenstates are given as 
\begin{equation}
|\varphi _1^{(S)}\rangle =\frac 1X(|LL\rangle -\frac{E_3}{\sqrt{2}w}%
|LR\rangle +|RR\rangle ),  \eqnum{15a}
\end{equation}
\begin{equation}
|\varphi _2^{(A)}\rangle =\frac 1{\sqrt{2}}(|RR\rangle -|LL\rangle ), 
\eqnum{15b}
\end{equation}
\begin{equation}
|\varphi _3^{(S)}\rangle =\frac 1Y(|LL\rangle -\frac a{\sqrt{2}w}|LR\rangle
+|RR\rangle ),\text{ \ \ }  \eqnum{15c}
\end{equation}
where we have defined the normalization constants $X=\sqrt{4w^2+E_3^2}/\sqrt{%
2}w$ and $Y=\sqrt{4w^2+E_1^2}/\sqrt{2}w$. The superscript $S$ ($A$) on the
left sides of Eq. (15) denotes symmetry (antisymmetry) under the spatial
reflection. We can see from Eq. (15) that due to strong Coulomb repulsion,
the symmetric ground state is dominated by the delocalized state $|LR\rangle 
$, whereas, the other two eigenstates are nearly degenerate and dominated by
the two localized states $|LL\rangle $ and $|RR\rangle $. Note that although 
$|\varphi _2^{(A)}\rangle $ and $|\varphi _3^{(S)}\rangle $ look like a
doublet in a single-electron double-trap system which consists of a pair of
symmetric and antisymmetric single-particle states, there are fundamental
differences for the present two-particle system. In fact, the superposition
of the two localized states $|LL\rangle $ and $|RR\rangle $ implies that the
spatial wave functions of the two electrons have been entangled, in the
usual sense that they are not factorized into single-particle states. To
describe the degree of entanglement, we define the maximally entangled Bell
state 
\begin{equation}
|\Psi _{Bell}\rangle =\frac 1{\sqrt{2}}(|RR\rangle +e^{i\phi }|LL\rangle ), 
\eqnum{16}
\end{equation}
with arbitrary phase angle $\phi $. Therefore the probability $\rho _{Bell}$
for finding the maximally entangled Bell state in a coupled quantum dot is
given by

\begin{equation}
\rho _{Bell}=\frac 12\left| C_3(t)+e^{i\phi }C_1(t)\right| ^2.  \eqnum{17}
\end{equation}
In the following we will investigate the various energy spectrum and
consequent dynamics for a GaAs ($m=0.067m_e$, $\kappa =13.1$) system
comprising two equal dots with vertical confinement energy $\omega _z=16$meV
($a_B=36$nm) and lateral confinement energy $\alpha \omega _z=8$meV. The
inter-dot distance is chosen to be $a=20$nm. The corresponding system
parameters are calculated to be $u=5.6$meV, and $w=-0.15$meV, implying $%
|\varphi _1^{(S)}\rangle \simeq |LR\rangle $, and $|\varphi _3^{(S)}\rangle
\simeq (1/\sqrt{2})(|RR\rangle +|LL\rangle )$.

\section{Localization preparation from delocalized ground state}

We start the search for localization with the simplest case, a constant
electric field $F(t)=F_0$. Before $t=0$ the system is in the delocalized
ground state $\mid \varphi _1^{(S)}\rangle $, and at $t=0$ the field $F_0$
is switched on suddenly. We find that when the value of electric field
satisfies the following resonance condition

\begin{equation}
F_0=u,  \eqnum{18}
\end{equation}
the initial delocalized ground state may be excited into a localized state
with two electrons occupying the same right dot. To elucidate this resonance
property, we show in Fig. 2 time evolution of the probabilities $%
P_{LR}(t)=\left| C_2(t)\right| ^2$ to find the two electrons in the
different dots (solid line), $P_{LL}(t)=\left| C_1(t)\right| ^2$ to find the
two electrons in the left dot (dashed line), and $P_{RR}(t)=\left|
C_3(t)\right| ^2$ to find the electrons in the right dot (dotted line). It
shows in Fig. 3 that both $P_{LR}(t)$ and $P_{RR}(t)$ oscillate between 0
and 1 with a definite period, while $P_{LL}(t)$ always negligibly small
during time evolution, meaning that a complete resonance takes place between
the delocalized state $\mid LR\rangle $ and localized state $\mid RR\rangle $%
. In this case, because the population of the localized state $\mid
LL\rangle $ in the left dot remains almost zero during time evolution, we
can neglect its contribution and describe the dynamics by the reduced
Schr\"{o}dinger equation

\begin{equation}
i\frac d{dt}\left( 
\begin{array}{l}
C_2(t) \\ 
C_3(t)
\end{array}
\right) =\left( 
\begin{array}{ll}
0 & \sqrt{2}w \\ 
\sqrt{2}w & 0
\end{array}
\right) \left( 
\begin{array}{l}
C_2(t) \\ 
C_3(t)
\end{array}
\right) .  \eqnum{19}
\end{equation}
Thus with the initial state $\mid \Psi (0)\rangle =\mid \varphi
_1^{(S)}\rangle \simeq \mid LR\rangle $, we have the evolution of the system
as follows 
\begin{equation}
C_2(t)=\cos (\sqrt{2}wt),  \eqnum{20a}
\end{equation}
\begin{equation}
C_3(t)=\exp (i\pi /2)\sin (\sqrt{2}wt).  \eqnum{20b}
\end{equation}
Clearly our two-state approximation Eq. (20) describes the dynamics very
well when compared with the exact numerical result shown in Fig. 2, implying
complete Rabi oscillation between the localized state $\mid RR\rangle $ and
delocalized state $\mid LR\rangle $ with oscillation period $\pi /(\sqrt{2}%
|w|)$.

Once the two electrons are localized, they can be forced to stay localized
permanently by switching the field to another nonzero value [see Fig. 3(a)].
We show this effect in Fig. 3(b). This trivial way of keeping localization
in a two-electron double-dot system originates from the fact that, due to
the presence of constant electric field, the consequent energy mismatch
among the three two-particle states prohibit tunneling from the localized
state $\mid RR\rangle $ to the other states.

Another way to implement localized two-particle state is through
adiabatically varying the constant electric field, which induces a series of
avoided crossings in the energy spectrum. Consequently, the electron number
in the ground state will experience a series of Coulomb stairs. This
field-induced adiabatic localization is shown in Fig. 4, where Fig. 4(a)
plots the evolution of electron number in the right dot for the ground state
as a function of the constant electric field, and Fig. 4(b) plots the
corresponding energy spectrum. It reveals in Fig. 4 that on adiabatically
increasing the strength of constant electric field, a series of avoided
crossings develop in the energy spectrum. Consequently, quantum transition
occurs at these avoided crossings, resulting in a series of Coulomb stairs.
We notice that adiabatically increasing the electric field to a typical
value of 2.8kV/cm enables complete localization of the ground state.

\section{Entanglement of two electrons in the presence of constant electric
field}

As shown in Fig. 2, a resonant constant electric field can induce a complete
oscillation between the delocalized state $|LR\rangle $ and localized state $%
|RR\rangle $. If the constant electric field is turned off [see Fig. 5(a)]
at time when the two electrons are fully localized in the right dot, as
shown in Fig. 2(b), the strong Coulomb repulsion will induce the resonance
between the localized states $\mid RR\rangle $ and $\mid LL\rangle $ during
subsequent evolution, whereas the delocalized state $\mid LR\rangle $ is
inhibited to be occupied. This dark property of the delocalized state is
shown in Fig. 5(b). It reveals in Fig. 5(b) that the value of $P_{LR}$ is
almost zero during time evolution, suggesting that the two electrons are
never separated into different dots. While cycling from one dot to the
other, the two electrons are always correlated and entangled, and very
likely to be found in the same dot.

The entanglement between the two electrons illustrated in Fig. 5(b) can be
well described by a two-state approximation. Because the population of the
delocalized state $\mid LR\rangle $ remains very small after time $t_0=\pi
/(2\sqrt{2}|w|)$ shown in Fig. 4(b), we can approximate $C_2(t)$ ($t>t_0$)
in Eq. (13) to first order of $w/u$

\begin{equation}
C_2(t)=\frac{-\sqrt{2}w}u\exp (-iut)[C_1(t)+C_3(t)].  \eqnum{21}
\end{equation}
By introducing $C_2(t)$ from Eq. (21) in the Schr\"{o}dinger equation we
reduce the system to an effective two-level system. The reduced equation has
the form

\begin{equation}
i\frac d{dt}\left( 
\begin{array}{l}
C_1(t) \\ 
C_3(t)
\end{array}
\right) =\left( 
\begin{array}{ll}
u & -2w^2/u \\ 
-2w^2/u & u
\end{array}
\right) \left( 
\begin{array}{l}
C_1(t) \\ 
C_3(t)
\end{array}
\right) .  \eqnum{22}
\end{equation}
Thus with the initial state $\mid \Psi (t_0)\rangle =-i\mid RR\rangle $ [see
Eq. (20)], we have the following time evolution of the system 
\begin{equation}
C_1(t)=\exp (-iut)\sin (2w^2t/u),  \eqnum{23a}
\end{equation}
\begin{equation}
C_3(t)=-i\exp (-iut)\cos (2w^2t/u).  \eqnum{23b}
\end{equation}
Substituting Eq. (23) into Eq. (17) we have the probability to find the Bell
state $(\mid RR\rangle +e^{i\phi }\mid LL\rangle )/\sqrt{2}$ at time $t$

\begin{equation}
\rho _{Bell}(t)=\frac 12[1+\sin (\omega _rt)\cos (\phi +\pi /2)],  \eqnum{24}
\end{equation}
where $\omega _r=4w^2/\kappa $. In particular we can see from Eq. (23) that
the system's quantum state at time

\begin{equation}
\tau =\pi u/8w^2+t_0,  \eqnum{25}
\end{equation}
corresponds to a $\phi =-\pi /2$ maximally entangled Bell state $(\mid
RR\rangle -i\mid LL\rangle )/\sqrt{2}$.

We present in Fig. 5(c) (solid line) the probability for finding the
maximally entangled Bell state ($\phi =-\pi /2$) as a function of time. The
system parameters are the same as that used in Fig. 5(b). Clearly our
two-state approximation [Eq. (24)] describes the system's evolution very
well when compared with the exact numerical solution shown in Fig. 5(c),
implying that the system's quantum state at time $\tau $ corresponds to a
maximally entangled Bell state ($\phi =-\pi /2$). However, the degree of
entanglement degrades after time $\tau $, a consequence of the fact that the
state at time $\tau $ is not an eigenstate of the field-free Hamiltonian. So
a single pulse of constant electric field can not preserve the entanglement
in our system. Note that no maximally entangled Bell-state generation is
possible if the effective Coulomb interaction $\kappa $, along with the
electric field, is turned off, as shown in Fig. 5(c) (dotted line). This
implies essential role of the nonlinear Coulomb interaction in forming the
entanglement between the electrons.

\section{Entanglement of two electrons in the presence of oscillatory
electric field}

We turn now to discussion of the entanglement in the presence of a
sinusoidal field of the form $F(t)=F_1\sin \omega t$. Figure 6 illustrates
the spectral features by plotting the Floquet quasienergies as a function of
driving frequency $\omega $. The amplitude value of the field is chosen to
be $F_1=1.5$kV/cm. It shows in Fig. 6 that on adiabatically increasing the
value of $\omega $, the quasienergies $\varepsilon _1$ and $\varepsilon _3$
approach each other. Especially when the value of driving frequency satisfies

\begin{equation}
\omega =u,  \eqnum{26}
\end{equation}
an avoided crossing occurs in the quasienergy spectrum.

To elucidate the effect of the avoided crossing displayed in Fig. 6 on the
quantum mechanical behavior of the system, we examine the dynamics of the
system starting with unperturbed ground state, i.e., $|\Psi (0)\rangle =$ $%
|\varphi _1^{(S)}\rangle $. Figure 7(a) shows time evolution of the
probabilities $P_{LR}(t)$ (solid line), $P_{LL}(t)$ (dashed line), and $%
P_{RR}(t)$ (dotted line), With the system parameters corresponding to the
avoided crossing shown in Fig. 6. It reveals in Fig. 7(a) that the two
electrons oscillate between the delocalized state $|LR\rangle $ and two
localized states $|LL\rangle $ and $|RR\rangle $. The occupations of two
localized states are always the same and the oscillations are in-phase. The
maximum values of $P_{LL}(t)$ and $P_{RR}(t)$ are $0.5$, which corresponds
to zero occupation of state $|LR\rangle $. Figure 7(b) shows the probability 
$\rho _{Bell}$ to find the maximally entangled Bell state with $\phi =\pi $.
We can see from Fig. 7(b) that the degree of entanglement varies with time.
In particular when the occupation of delocalized state $|LR\rangle $ is
zero, the two electrons are maximally entangled with $\rho _{Bell}=1$.

Once the two electrons are in the maximally entangled Bell state, they can
remain maximally entangled by suddenly turning off the oscillatory electric
field [see Fig. 8(a)]. We show this effect in Figs. 8(b)-(c) where Fig. 8(b)
plots time evolution of the occupations of three two-particle states and
Fig. 8(c) the probability to find the maximally entangled Bell state with $%
\phi =\pi $ (solid line). It shows in Fig. 8(b) that a pulse of oscillatory
electric field induces the two electrons to stay on the same dot, while each
of them occupies either of the dots with the same probability. It reveals in
Fig. 8(c) (solid line) that the two electrons remain maximally entangled
during time evolution after the oscillatory field is turned off. This is
different from what is shown in Fig. 5(c) where the degree of entanglement
varies with time. Therefore, by a pulse of oscillatory electric field with
frequency satisfying the resonant condition Eq. (26), the maximally
entangled Bell state can be created and maintained. Loss and Divincenzo
studied entanglement in double quantum dots involving the spin degree of
freedom\cite{Loss}. Here we have identified a complementary method that
creates and preserves entanglement between the spatial wave functions of two
electrons in a coupled quantum dot.

The generation of maximally entangled Bell state shown in Fig. 8 can be well
described by a two-state approximation. Taking into account symmetric
properties of the three unperturbed eigenstates of the system, the dynamics
is determined by the one-photon transition between the ground state $%
|\varphi _1^{(S)}\rangle $ and the first excited two-particle state $%
|\varphi _2^{(A)}\rangle $, whereas the transition from $|\varphi
_1^{(S)}\rangle $ to $|\varphi _3^{(S)}\rangle $ is prohibited due to their
identical symmetry. In this case, we can approximate the Hamiltonian $H_1(t)$
in a Hilbert space spanned by the states $|\varphi _1^{(S)}\rangle $ and $%
|\varphi _2^{(A)}\rangle $

\begin{equation}
H_1(t)=\left( 
\begin{array}{cc}
E_1 & -\sqrt{2}\mu _1(t)/X \\ 
-\sqrt{2}\mu _1(t)/X & E_2
\end{array}
\right) .  \eqnum{27}
\end{equation}
In the interaction representation and after applying the rotating-wave
approximation (RWA) the Schr\"{o}dinger Equation takes the form

\begin{equation}
i\frac d{dt}\left( 
\begin{array}{c}
d_1 \\ 
d_2
\end{array}
\right) =\left( 
\begin{array}{cc}
0 & -\Omega _r \\ 
-\Omega _r & \Delta
\end{array}
\right) ,  \eqnum{28}
\end{equation}
where $d_1$ and $d_2$ are the probability amplitudes of the bare states $%
|\varphi _1^{(S)}\rangle $ and $|\varphi _2^{(A)}\rangle $, $\Omega _r=V_1/%
\sqrt{2}X$ is Rabi frequency, and $\Delta =E_2-E_1-\omega $ is the detuning
of the driving frequency $\omega $ from the transition frequency. In the
case of one-photon resonance $E_2-E_1=\omega $, we obtain the time evolution
of initial ground state ($d_1(0)=1$, $d_2(0)=0$) as follows 
\begin{equation}
d_1=\cos \Omega _rt,  \eqnum{29a}
\end{equation}
\begin{equation}
d_2=i\sin \Omega _rt.  \eqnum{29b}
\end{equation}
Thus the system oscillates between the ground state wherein two electrons
are highly delocalized and the first excited state wherein the two electrons
are highly entangled. Substituting Eq. (29) into Eq. (17), and in the weak
coupling limit $w\ll u$, we obtain the expression for the probability to
find the maximally entangled Bell state

\begin{equation}
\rho _{Bell}(t)=\frac 12(1-\cos \phi )\sin ^2(w\mu _1t/u),  \eqnum{30}
\end{equation}
where we have approximated Rabi frequency $\Omega _r$ with $w\mu _1/u$. In
particular we can see from Eq. (29) that the quantum state of the system at
time

\begin{equation}
\tau =\pi u/(2w\mu _1),  \eqnum{31}
\end{equation}
corresponds to a $\phi =\pi $ maximally entangled Bell state $(|RR\rangle
-|LL\rangle )/\sqrt{2}$.

The result of Eq. (30) is shown in Fig. 8(c) (dotted line). Clearly, in
comparison with the exact numerical solution, our two-state approximation
describes the system evolution very well, suggesting that the quantum state
of system at time $\tau =\pi u/(2w\mu _1)$ corresponds to a maximally
entangled Bell state of the desired form with $\phi =\pi $. Therefore we
arrive at the conclusion that a selective pulse of oscillatory electric
field with duration $\tau =\pi u/(2w\mu _1)$ can be used to create and
maintain a maximally entangled Bell state ($\phi =\pi $) in the system of
two electrons in a coupled quantum dot.

We notice from Eq. (31) that Bell-state generation time is significantly
shortened by increasing the amplitude of the oscillatory electric field.
This is important because shorter Bell-state generation time is fundamental
to the experimental observation of such maximally entangled state, which is
impeded by inevitable decoherence occurred in the realistic double quantum
dot system. The decoherence is the most problematic issue pertaining to most
quantum computing processing. In the present entangled state proposal, the
decohering time depends partly on the fluctuation of the single particle
energy caused by the modification of the confining potential due to phononic
excitations. There is also a quantum electrodynamic contribution because of
coupling to the vacuum modes. In addition, impurity scattering and phonon
emission also have contributions to the decoherence. However, in principle,
their effects can be minimized by more precise fabrication technology and by
cooling the system.

\section{Dynamical localization of two interacting electrons}

In section IV we have shown that a trivial way of maintaining localization
is to suddenly shift the constant field to another value once the electrons
are in the localized state $|RR\rangle $. In this section we study the
possibility of remaining the localization with the oscillatory electric
field $F(t)=F_1\sin \omega t$\cite{Ping3}. Because the localized state $%
|RR\rangle $ may be always produced from the ground state as described
above, we therefore suppose in the following discussions that the system
starts with the localized state $|RR\rangle $.

In the presence of time-dependent electric field, the evolution of the
system can not be solved in a closed form because $[H_1(t_1),H_1(t_2)]\neq 0$%
. However, time periodicity of the Hamiltonian (11) enables us to describe
the dynamics within the Floquet formalism. We numerically integrate the
motion of equation for time evolution operator

\begin{equation}
i\frac \partial {\partial t}U(t,0)=H_1(t)U(t,0),  \eqnum{32}
\end{equation}
and diagonalize $U(2\pi /\omega ,0)$ to obtain the quasienergies $%
\{\varepsilon _{\alpha ,l}\}$ and Floquet states $\{\mid u_{\alpha
,l}(0)\rangle \}$ at time $t=0$. Here the quasienergies $\varepsilon
_{\alpha ,l}$ are confined to the first Brillouin zone and, at $F_1=0$,
connected to $E_\alpha +l\omega $. The index $l$ counts `how many photons'
have to be subtracted from the unperturbed energy level $E_\alpha $ in order
to arrive in the first Brillouin zone. The Floquet state $\mid u_{\alpha
,l}(t)\rangle $ can be obtained from the eigenvalue equation

\begin{equation}
(H_1(t)-i\frac \partial {\partial t})\mid u_{\alpha ,l}(t)\rangle
=\varepsilon _{\alpha ,l}\mid u_{\alpha ,l}(t)\rangle ,  \eqnum{33}
\end{equation}
where $\alpha =1,2,3$. Note that the Hamiltonian (11) remains invariant
under the combined spatial reflection and time translation $t\rightarrow
t+\pi /\omega $. An immediate consequence of this dynamical symmetry is that
each Floquet state is either odd or even\cite{Peres,Shin}. When the driving
amplitude is switched off adiabatically, $F_1\rightarrow 0$, the Floquet
states are connected with the stationary eigenstates in Eq. (15) as follows 
\cite{Grifoni}

\begin{equation}
\mid u_{\alpha ,l}(t)\rangle \rightarrow \mid u_{\alpha ,l}^0(t)\rangle
=\varphi _\alpha exp(il\omega t).  \eqnum{34}
\end{equation}
Thus we can easily determine the dynamical parity of the Floquet state $\mid
u_{\alpha ,l}(t)\rangle $.\ 

We present in Figs. 9(a)-(b) the quasienergies versus the amplitude $F_1$,
where the values of driving frequency $\omega $ are respectively chosen to
satisfy $u=3.1\omega $ and $u=2\omega $, respectively. In Fig. 9(a) we see
that the quasienergies $\varepsilon _{2,-3}$ and $\varepsilon _{3,-3}$ with
different parity form an exact crossing at $F_1=1.05$kV/cm. When $%
F_1\rightarrow 0$, all three quasienergies arrive at the unperturbed cases
by $\varepsilon _{1,0}^0=E_1$, $\varepsilon _{2,-2}^0=E_2-3\omega $, $%
\varepsilon _{3,-2}^0=E_3-3\omega $. For $2l\omega \rightarrow $ $E_2-E_1$, $%
\varepsilon _{1,0}$ and $\varepsilon _{3,-2l}$, belonging to different
parity, are also allowed to cross, as shown in Fig. 9(b) where the value of
index is $l=1$. In the exact same way, when $(2l+1)\omega \rightarrow $ $%
E_3-E_1$ two quasienergies $\varepsilon _{1,0}$ and $\varepsilon
_{3,-(2l+1)} $ can develop into the crossing for special value of the
amplitude $F_1$ (not shown here).

To elucidate the effect of the exact crossing on the quantum mechanic
behavior of the system we present in Fig. 10(a) the time evolution of $%
P_{RR}(t)$ with the system parameters corresponding to the first crossing
between $\varepsilon _{2,-3}$\ and $\varepsilon _{3,-3}$ shown in Fig. 9(a).
For comparison we also plot $P_{RR}(t)$ in Fig. 10(b) for the value of
amplitude $F_1=0.5$kV/cm. In Fig. 10(a) we can see that during time
development the probability $P_{RR}(t)$ remains near 1 as if the two
electrons were frozen in the right dot. Thus at the exact crossing of $%
\varepsilon _{2,-3}$\ and $\varepsilon _{3,-3}$ the dynamical localization
builds up although the strong Coulomb repulsion between the two electrons
prevents the system behaving so. Note that the value of $u$ used in Fig.
10(a) corresponds to the unperturbed energies $E_1=-0.016$meV, $E_2=5.6$meV,
and $E_3=5.616$meV. Because $E_1$ is much lower than $E_2$ and $E_3$, the
unperturbed eigenstates $\varphi _2^A$ and $\varphi _3^S$, both of which
have a very small component of the delocalized two-particle state $\mid
LR\rangle $, become comparable to a doublet in a symmetrical double-dot
system. So it is expected that at the crossing of the quasienergies $%
\varepsilon _{2,-3}$\ and $\varepsilon _{3,-3}$ the initial localized state
can be approximated by a superposition of \ degenerate Floquet states $\mid
u_{2,-3}(0)\rangle $ and $\mid u_{3,-3}(0)\rangle $, which remains localized
in perpetuity. This looks like the case of a single-electron, two-level
system consisting of the lowest symmetric and antisymmetric states of the
double-trap potential, where perfect localization can be achieved at the
exact crossing between the two Floquet quasienergies. It is numerically
found that even if the Coulomb interaction $u$ is strong enough, the
dynamical localization can still occur so long as the quasienergies $%
\varepsilon _{2,m}$ and $\varepsilon _{3,m}$ cross each other. Moreover, the
value of the system parameter $2\mu _1/\omega $\ corresponding to the first
crossing is about $2.4$, which is a\ root of the zero-order Bessel function,
suggesting that in this situation the dynamical localization can be
approximated by the driven two-level model. If the system parameters deviate
from the level crossing, then the dynamical localization ceases to exist and 
$P_{RR}$ oscillates between $0$ and $1$ in the time development,\ as shown
in Fig. 10(b).

More surprisingly, we notice that compared with non-interacting ($u=0$)
case, the presence of Coulomb repulsion enhances the degree of localization
to a larger extent. To illustrate this feature we present in Fig. 10(c) time
evolution of $P_{RR}$ with the parameters corresponding to the first exact
crossing of the quasienergies in the absence of Coulomb interaction. It
shows in Fig. 10(c) that although dynamical localization maintains during
time development, the degree of localization is much lower than that shown
in Fig. 10(a), suggesting Coulomb-enhanced localization. From Eq. (15) we
know that due to strong Coulomb interaction, the initial localized state $%
|RR\rangle $ can be approximated by a superposition of two eigenstates $%
|\varphi _2^{(A)}\rangle $ and $|\varphi _3^{(S)}\rangle $ as $|RR\rangle
\simeq (1/\sqrt{2})(|\varphi _2^{(A)}\rangle +|\varphi _3^{(S)}\rangle )$.
Thus from what we have learned in single-particle double-trap system, it is
not difficult to understand why the localization can be dynamically remained
at the exact crossing of two Floquet states developed from $|\varphi
_2^{(A)}\rangle $ and $|\varphi _3^{(S)}\rangle $. Similarly, in the absence
of Coulomb interaction, the initial localized state $|RR\rangle $ can be
approximated by a superposition of all three eigenstates as $|RR\rangle =(1/%
\sqrt{2})(|\varphi _1^{(S)}\rangle +|\varphi _2^{(A)}\rangle )+|\varphi
_3^{(S)}\rangle $, suggesting dynamical localization at the crossing among
three quasienergies. However, the fundamental difference lies in the fact
that when the strong interaction is present, the tunneling coupling is $%
\langle LL|H_1|RR\rangle =E_3-E_2\simeq 4w^2/u$. In the absence of Coulomb
interaction, whereas, the tunneling coupling is $\langle LL|H_1|RR\rangle
=(1/2)(E_1+E_3)-E_2=2w$. Thus we can see that the coupling between the two
localized states greatly decreases in the presence of strong Coulomb
interaction, which leads to fundamental increase of localization degree.

We turn to study the dynamics of the system at the crossing of the
quasienergies $\varepsilon _{1,0}$ and $\varepsilon _{2,-2}$ in Fig. 9(b),
using the same initial state condition. The result is shown in Fig. 10(d).
In contrary to that shown in Fig. 10(a), it shows in Fig. 10(d) that at the
level crossing of $\varepsilon _{1,0}$\ and $\varepsilon _{2,-2}$ dynamical
localization does not happen and $P_{RR}(t)$ oscillates between $0$ and $1$.
Note that the level crossing of $\varepsilon _{1,0}$\ and $\varepsilon
_{2,-2}$ induces strong participation of the Floquet state $\mid
u_{1,0}(t)\rangle $ during the time evolution of the system, and the most
component in $\mid u_{1,0}(t)\rangle $ is the delocalized two-particle state 
$\mid LR\rangle $. Therefore the strong mixture of the Floquet state $\mid
u_{1,0}(t)\rangle $ in the evolution of the system will lead to complete
destruction of the dynamical localization, as shown in Fig. 10(d).

In the above discussions we have ignored higher-lying single-particle
states; this requires that the frequency of the external field is much lower
than single-particle level spacing. In the presence of decoherence due to
environmental dissipation, long dephasing time should be required. A
detailed analysis of the effect of a decohering environment will be given
elsewhere.

\section{Conclusions}

In conclusion we have shown how dynamical localization and entanglement of
two interacting electrons in a double quantum dot system can be accessed by
external electric fields. We have found that (i) The presence of a constant
electric field may induce the complete Rabi oscillation between the
delocalized state $\mid LR\rangle $ and localized state $\mid RR\rangle $.
Thus Starting from the delocalized ground state, we can prepare a fully
localized state. The localization can be maintained by switching the field
to another nonzero value. (ii) The two electrons oscillate between the
delocalized state and two localized states in the presence of a resonant
oscillatory field. With the oscillatory field turned off at time when the
probabilities for finding the electrons in the left and right dot are
identically $0.5$, the two electrons remain maximally entangled in the
subsequent time evolution. Thus a selective pulse of oscillatory field can
be used to implement maximally entangled Bell states in a two-electron
two-dot system. (iii) Although the Coulomb repulsion is very strong, the two
initially localized electrons can stay localized during time evolution. It
is also shown that compared to non-interacting case, the Coulomb interaction
enhances the degree of localization to a larger extent. We expect the
present results are useful in exploiting the coherent control of electrons
in quantum dot systems.

\begin{center}
{\bf ACKNOWLEDGMENTS}
\end{center}

This work was supported partly by the National Natural Science Foundations
of China under Grant No. .

{\Large Figure captions}

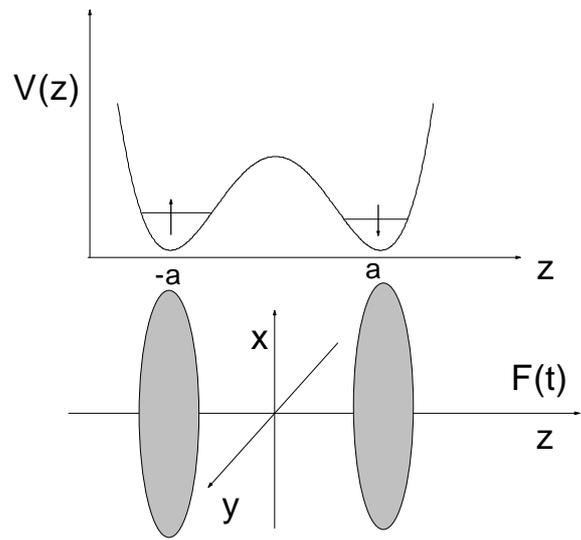
{\bf Fig. 1 }Sketch{\bf \ }of two interacting elctrons in a coupled quantum
dot driven by electric fields.{\bf \ }

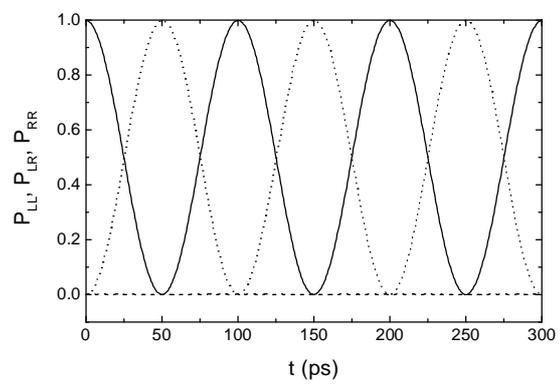
{\bf Fig.2. }Time evolution of the probabilities $P_{LR}$ (solid line), $%
P_{LL}$ (dashed line), and $P_{RR}$ (dotted line) for the value of the
strength of constant electric field satisfying resonant condition $\mu _0=u$.

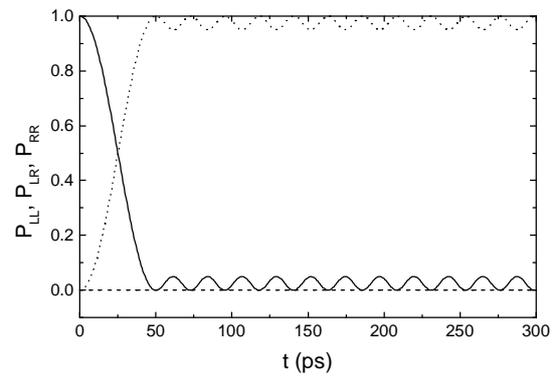
{\bf Fig. 3. }(a){\bf \ }Electric field that imposes on the double quantum
dot system; (b) Time evolution of the probabilities $P_{LR}$ (solid line)
and $P_{RR}$ (dotted line) under the influence of the electric field shown
in (a). The other system parameters are the same as that used in Fig. 2.

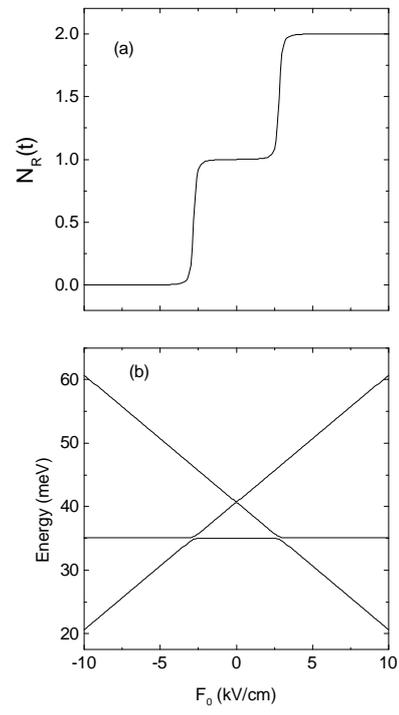
{\bf Fig. 4. }(a) Electron number distribution of right dot in the ground
state as a function of the strength of constant electic field; (b) Energy
spectrum of the driven two-electron system as a function of the strength of
a constant electric field.

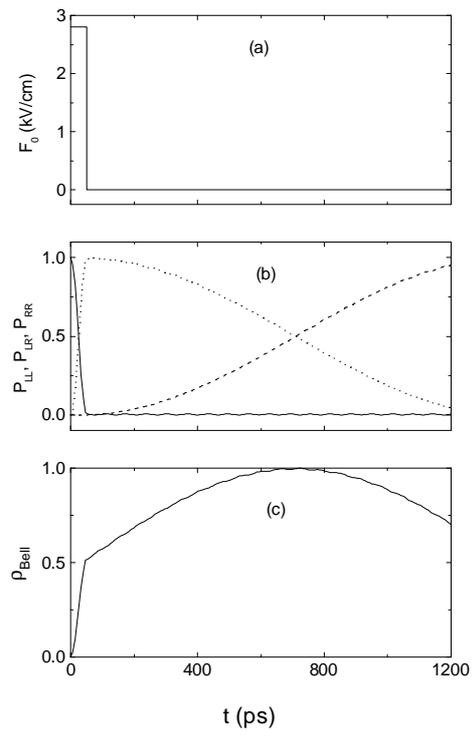
{\bf Fig. 5. }(a){\bf \ }Electric field that imposes on the double quantum
dot system; (b) Time evolution of the probabilities $P_{LR}$ (solid line)
and $P_{RR}$ (dotted line) under the influence of the electric field shown
in (a); (c) Time evolution of the probability $\rho _{Bell}$ for finding the
maximally entangled Bell state ($\phi =-\pi /2$) in the presence of the
Coulomb interaction (solid line) and in the absence of the Coulomb
interaction (dotted line).

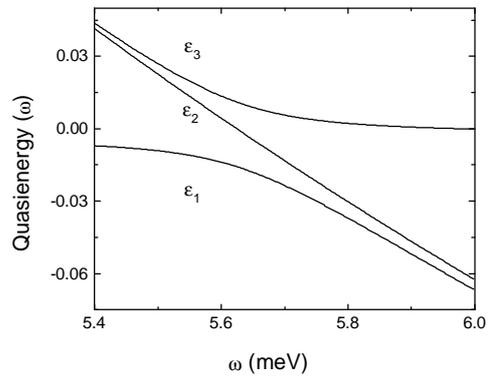
{\bf Fig. 6. }Quasienergy spectrum of the driven two-electron system as a
function of the frequency of oscillatory electric field.

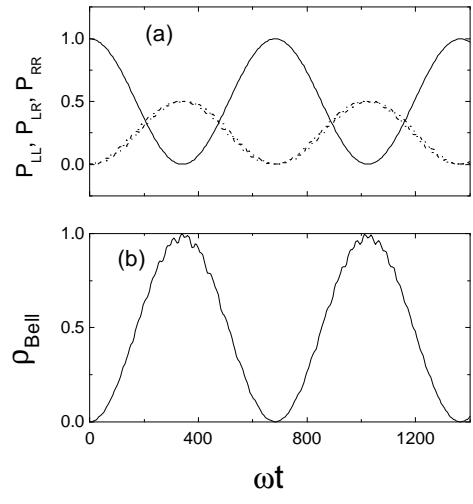
{\bf Fig. 7. }(a) Time evolution of the probabilities $P_{LR}$ (solid line), 
$P_{LL}$ (dashed line), and $P_{RR}$ (dotted line) for the value of the
frequency of oscillatory electric field $\omega =u$; (b) Time evolution of
the probability $\rho _{Bell}$ for finding the maximally entangled Bell
state ($\phi =\pi $), The system parameters are the same as that used in (a).

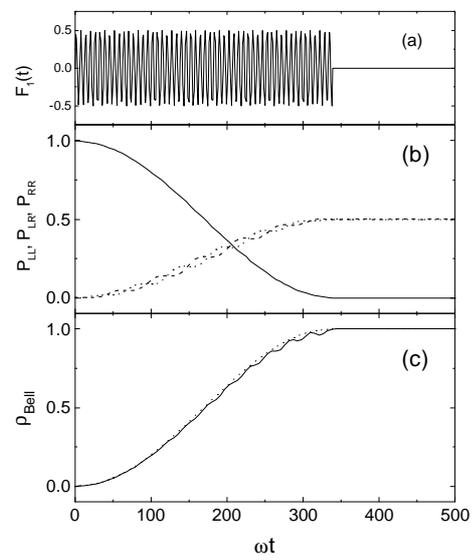
{\bf Fig. 8. }(a){\bf \ }Electric field that imposes on the double quantum
dot system; (b) Time evolution of the probabilities $P_{LR}$ (solid line), $%
P_{LL}$ (dashed line), and $P_{RR}$ (dotted line) under the influence of the
electric field shown in (a); (c) Exact numerical (solid line) and
approximate analytic (dotted) results of time evolution of the probability $%
\rho _{Bell}$ for finding the maximally entangled Bell state ($\phi =\pi $)
in a coupled quantum dot. The other system parameters are the same as that
used in Fig. 7.

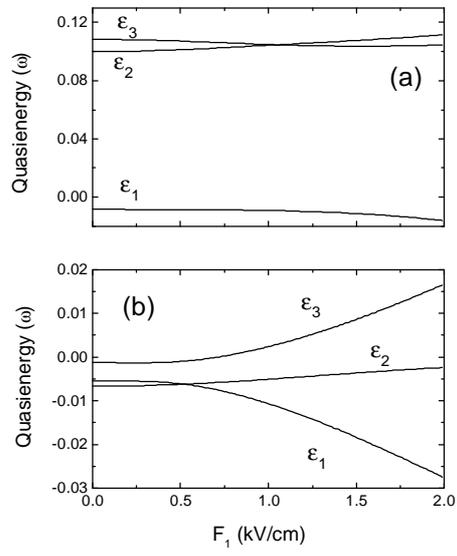
{\bf Fig. 9. }Floquet spectrum of the driven two-electron system as a
function of the strength of oscillatory electric field for the value of the
effective Coulomb interaction (a) $u=3.1\omega $; (b) $u=2\omega $.

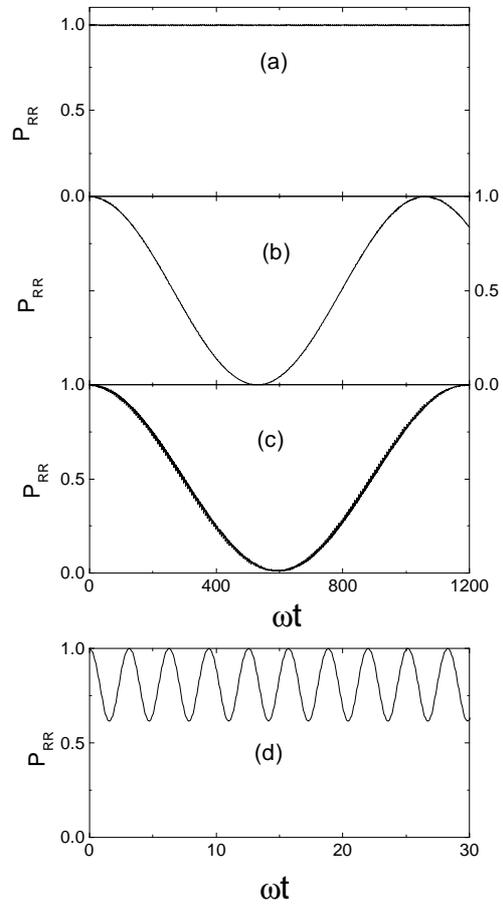
{\bf Fig. 10. }Time evolution of the probability $P_{RR}(t)$ for four
different kinds system parameter values (a) $u=3.1\omega $ and $F_1=1.05$%
kV/cm, corresponding to the exact level crossing shown in Fig. 9(a); (b) $%
u=3.1\omega $ and $F_1=0.5$kV/cm, for comparison with the case shown in (a);
(c) $u=0$, $\omega =1.81$meV, and $F_1=2$kV/cm, corresponding to the exact
level crossing in the absence of Coulomb interaction; (d) $u=2\omega $ and $%
F_1=0.52$kV/cm, corresponding to the exact level crossing shown in Fig. 9(b).


\begin{references}
\bibitem{Grifoni}  M. Grifoni and P. H\"{a}nggi, Phys. Rep. {\bf 304}, 229
(1998).

\bibitem{Gross1}  F. Grossmann, T. Dittrich, P. Jung, and P. H\"{a}nggi,
Phys. Rev. Lett. {\bf 67}, 516 (1991).

\bibitem{Bavli}  R. Bavli and H. Metiu, Phys. Rev. Lett.{\bf \ 69}, 1986
(1992).

\bibitem{Gomez}  J. M. Gomez Llorente and J. Plata, Phys. Rev. A {\bf 45},
R6958 (1992).

\bibitem{Gross}  F. Grossmann and P. H\"{a}nggi, Europhys. Lett. {\bf 18},
571 (1992).

\bibitem{Hone}  M. Holthous and D. Hone, Phys. Rev. B {\bf 47}, 6499 (1993).

\bibitem{Cuk}  R.I. Cukier and M. Morillo, Chem. Phys. {\bf 183}, 375 (1994).

\bibitem{Zuo}  T. Zuo, S. Chekowski, and A.D. Bandrauk, Phys. Rev. A {\bf 49}%
, 3943 (1994).

\bibitem{Pu}  H. Pu, S. Raghavan, and N. P. Bigelow, Phys. Rev. A\ {\bf 61},
023602 (2000).

\bibitem{ping}  P. Zhang, A.Z. Zhang, S.Q. Duan, and X.-G. Zhao, Phys. Lett.
A 287, 105 (2001).

\bibitem{Ping2}  P. Zhang, Q.-K. Xue, X.-G. Zhao, and X.C. Xie, Phys. Rev. A 
{\bf 66}, 022117 (2002).

\bibitem{Metiu}  P.I. Tamborenea and H. Metiu, Europhys. Lett. {\bf 53}, 776
(2001).

\bibitem{Steane}  A. Steane, Rep. Prog. Phys. {\bf 61}, 117 (1998).

\bibitem{Tur}  Q.A. Turchette, C.S. Wood, B.E. King, C.J. Myatt, D.
Leibfried, W.M. Itano, C. Monroe, and D.J. Wineland, Phys. Rev. Lett. {\bf 81%
}, 3631 (1998).

\bibitem{Hagley}  E. Hagley, X. Maitre, G. Nogues, C. Wunderlich, M. Brune,
J.-M. Raimond, and S. Haroche, Phys. Rev. Lett. {\bf 79}, 1 (1997).

\bibitem{Rau}  A. Rauschenbeutel, G. Nogues, S. Osnaghi, P. Bertet, M.
Brune, J.-M. Raimond, and S. Haroche, Science {\bf 288}, 2024 (2000).

\bibitem{Sack}  C.A. Sackett, D. Kielpinski, B.E. King, C. Langer, V. Meyer,
C.J. Myatt, M. Rowe, Q.A. Turchette, W.M. Itano, D.J. Wineland, and C.
Monroe, Nature {\bf 404}, 256 (2000).

\bibitem{Oo}  F. R. Waugh, M. J. Berry, D. J. Mar, R. M. Westvelt, K. L.
Campman, and A. C. Gossard, Phys. Rev. Lett. {\bf 75}, 705 (1995); C.
Livermore, C. H. Crouch, R. M. Westvelt, K. L. Campman, and A. C. Gossard,
Science {\bf 274}, 1332 (1996); R. H. Blick, D. Pfannkuche, R. J. Haug, K.
v. Klitzing, and K. Eberl, Phys. Rev. Lett. {\bf 80}, 4032 (1998); R. H.
Blick, D. W. vander Weide, R. J. Haug, and K. Eberl, {\it ibid. }{\bf 81},
689 (1998). See also references in these publications.

\bibitem{To}  For recent articles on double quantum dots with few electrons
see: Y. Tokura, D.G. Austing, and S. Tarucha, J. Phys.: Condens. Matter {\bf %
11,} 6023 (1999); B. Partoens, A. Matulis, and F.M. Peeters, Phys. Rev. B 
{\bf 59,} 1617 (1999); D.G. Austing, T. Honda, K. Muraki, Y. Tokura, and S.
Tarucha, Physica B {\bf 249}, 206 (1998); and references therein.

\bibitem{Kane}  B.E. Kane, Nature{\bf \ 393}, 133 (1998).

\bibitem{Loss}  D. Loss, and D.P. DiVincenzo, Phys. Rev. A {\bf 57}, 120
(1998).

\bibitem{Ping3}  P. Zhang and X.-G. Zhao, Phys. Lett. A {\bf 271}, 419
(2000).

\bibitem{Imamoglu}  A. Imamoglu, D.D. Awschalom, G. Burkard, D.P.
DiVincenzo, D. Loss, M. Sherwin, and A. Small, Phys. Rev. Lett. {\bf 83},
4204 (1999).

\bibitem{Quiroga}  L. Quiroga and N.F. Johnson, Phys. Rev. Lett. {\bf 83},
2270 (1999).

\bibitem{Jacak}  L. Jacak, P. Hawrylak, and A. W\'{o}js, {\it Quantum dots }%
(Springer, Berlin, 1997).

\bibitem{Burcard}  G. Burcard, G. Seelig, and D. Loss, preprint
(cond-mat/9910105).

\bibitem{Schliemann}  J. Schliemann, D. Loss, and A.H. MacDonald, preprint
(cond-mat/0009083).

\bibitem{Peres}  A. Peres, Phys. Rev. Lett. {\bf 67}, 158 (1991).

\bibitem{Shin}  J. Y. Shin and H. W. Lee, Phys. Rev. E {\bf 53}, 3096 (1996).
\end{references}
\end{document}